\documentclass{sig-alternate}

\usepackage{xcolor}
\usepackage{ulem}
\usepackage{enumerate}
\usepackage{etoolbox}

\usepackage[shortlabels]{enumitem}
\setlist[enumerate,1]{leftmargin=0.4cm}

\usepackage{pdfcomment}

\usepackage{eso-pic}

\usepackage{cleveref}
\crefname{paragraph}{\S}{\S\S} % default is {paragraph}{paragraphs}

\newcommand{\suppress}[1]{}  %%ignore its one argument, used to leave legacy text until we're sure e don't want it.
\newcommand{\chpc}{\textsf{cHPC}}
\AddToShipoutPicture{%
           \AtPageCenter{%
        \put(120,-500){\rotatebox[origin=c]{45}{\parbox[t]{10cm}{\color[gray]{0.80}%
        \scalebox{14}{DRAFT}\par
    }}}}%
}%i

\makeatletter
\def\@copyrightspace{\relax}
\makeatother

\begin{document}

\title{Rethinking High Performance Computing Platforms: Challenges, Opportunities and Recommendations}

\numberofauthors{4}

\author{
% 1st. author
\alignauthor Ole Weidner\\
    \affaddr{School of Informatics}\\
    \affaddr{University of Edinburgh, UK}\\
    \email{ole.weidner@ed.ac.uk}
% 2nd. author
\alignauthor Malcolm Atkinson \\
    \affaddr{School of Informatics}\\
    \affaddr{University of Edinburgh, UK}\\
    \email{malcolm.atkinson@ed.ac.uk}
\and
% 3th. author
\alignauthor Adam Barker\\
    \affaddr{School of Computer Science}\\
    \affaddr{University of St Andrews, UK}\\
    \email{adam.barker@st-andrews.ac.uk}
% 4rd. author
\alignauthor Rosa Filgueira\\
    \affaddr{School of Informatics}\\
    \affaddr{University of Edinburgh, UK}\\
    \email{rosa.filgueira@ed.ac.uk}
}

\date{01 February 2016}

\maketitle
\begin{abstract}

A new class of ``Second generation'' high-performance computing applications with
heterogeneous, dynamic and data-intensive properties have an extended set of requirements, which
cover application deployment, resource allocation, -control, and I/O scheduling. These requirements
are not met by the current production HPC platform models and policies. This results in a loss of
opportunity, productivity and innovation for new computational methods and tools. It also decreases
effective system utilization for platform providers due to unsupervised workarounds and ``rogue''
resource management strategies implemented in application space. In this paper we critically discuss
the dominant  HPC platform model and describe the challenges it creates for second generation
applications because of its \textit{asymmetric} resource view, interfaces and software deployment
policies. We present an extended, more \textit{symmetric} and application-centric platform model
that adds decentralized deployment, introspection, bidirectional control and information flow and
more comprehensive resource scheduling. We describe \textit{cHPC}: an early prototype of a
non-disruptive implementation based on Linux Containers (LXC). It can operate alongside existing
batch queuing systems and exposes a symmetric platform API without interfering with existing
applications and usage modes. We see our approach as a viable, incremental next step in HPC platform
evolution that benefits applications and platform providers alike. To demonstrate this further, we
layout out a roadmap for future research and experimental evaluation.

\end{abstract}




%
%  Use this command to print the description
%

% We no longer use \terms command


%%%%%%%%%%%%%%%%%%%%%%%%%%%%%%%%%%%%%%%%%%%%%%%%%%%%%%%%%%%%%%%%%%%%%%%%%%%%%%%%%%%%%%%%%%%%%%%%%
%%
\section{Introduction}
\label{sec:intro}

With computational methods, tools and workflows becoming ubiquitous in more and more scientific
domains and disciplines, the software applications and user communities on high performance
computing platforms are rapidly growing diverse. Many of the emerging second generation HPC
applications move beyond tightly-coupled, compute-centric methods and algorithms and embrace more
heterogeneous, multi-component workflows, dynamic and ad-hoc computation and data-centric
methodologies. While diverging from the traditional HPC application profile, many of these
applications still rely on the large number of tightly coupled cores, cutting-edge hardware and
advanced interconnect topologies provided only by HPC clusters. Consequently, HPC platform providers
often find themselves faced with requirements and requests that are so diverse and dynamic that they
become increasingly difficult to fulfill efficiently within the current operational policies and
platform models. The balancing act between supporting stable production science on the one hand and
novel application types and exploratory research in high-performance, distributed, and scientific
computing on the other, puts an additional strain on platform providers. In many places this
inevitably creates dissatisfaction and friction between the platform operators and their user
communities. However, this largely remains an unquantified, subjective perception throughout the
user and platform provider communities. We believe that platform providers and users have a common
mission to push the edge of the envelope of scientific discovery. Friction and dissatisfaction
creates an unnecessary loss of momentum and in the worst cases can cause productivity and innovation
to stall.

Cloud computing offers an alternative paradigm to traditional HPC platforms. Clouds offer on-demand
utility computing as a service, with resource elasticity and pay as you go pricing. One important
aspect to address is whether our vision of a symmetric HPC platform isn't just trying to turn HPC
platforms into \textit{Cloud-like} environments. This depends on the class and structure of the
application. Loosely coupled, elastic applications, which don't require guaranteed performance make
good candidates for virtualized environments. However, there are several classes of application
which typically run more effectively on dedicated HPC platforms: (1) Performance sensitive
applications: virtualization offers a significant performance overhead and due to multi tenancy,
application performance cannot usually be guaranteed; (2) Interconnect-sensitive applications, which
require co-location low latency and high throughput; (3) I/O-sensitive applications that, without a
very fast I/O subsystem, will run slowly because of storage bottlenecks; (4) Applications which
require dedicated and specialized hardware to run the computation; (5) Finally (an aspect which is
often overlooked) is the fact that the cost of migrating and storing data in the cloud is high. Data
intensive or big data applications often have data which is tethered to a site and therefore the
only viable option is to run the application on dedicated institutional resources.

Our primary objective will deliver HPC platforms that provide more flexible mechanisms and
interfaces for applications that are inherently dependent on their architectural advantages.
Otherwise, we fear that evolution and innovation in second generation applications might come to a
grinding halt as the platform is too confining for advanced use-cases. We argue that these confining
issues are largely caused by a \textit{structural asymmetry} between platforms and applications.
This asymmetry can be observed in the operational policies and as a consequence in production HPC
platform models and their technical implementations. Operational policies are characterized by a
centralized \textit{software deployment processes} managed and executed by the platform operators.
This impedes application development and deployment by creating a central bottleneck, especially for
second generation applications that are built on non-standard or even experimental software.
Production HPC platform models are characterized by \textit{static resource mapping} , an
\textit{asymmetric resource model}, and \textit{limited information exchange and control channels}
between platforms and their tenants. In this paper we propose changes to policies and platform
models to improve the application context without jeopardizing platform stability and reliability:

\noindent
\begin{enumerate}

  \setlength{\itemsep}{0pt}
  \setlength{\parskip}{1pt}
  \setlength{\parsep}{1pt}

\item By moving away from deployment monopolies, software provisioning can be handled directly by
the users and application experts, reducing bottlenecks, supporting increased application mobility,
and creating a shared sense of responsibility and a better balance between the two stakeholders.
This allows platform operators to focus on running HPC platforms at optimal performance,
reliability, and utilization.
%It will reduce waste from application runs that fail to produce
%results due to platform-related factors.

\item A more symmetric resource model, information and control flow between platform and application
will significantly improve platform operation while supporting the development and adoption of
innovative applications, higher-level frameworks and supporting libraries. We show via practical
examples and existing research how a platform model that is built upon more symmetric information
and control flow can provide a solid supporting foundation for this. The more applications will
exploit this foundation, the easier it becomes to operate an HPC platform at the desired optimal
point of performance, reliability, and utilization. We suggest how to amend and extend existing
platform models and their implementations.

\end{enumerate}

\noindent This paper is structured as follows: In \cref{sec:focus_applications} we present our
experience and observations from working with three different classes of second generation
applications on production HPC platforms: \textit{dynamic applications}
(\cref{sec:dynamic_applications}), \textit{data-intensive applications}
(\cref{sec:data_intensive_applications}), and \textit{federated applications}
(\cref{sec:ds_research_prototypes}). In \cref{sec:policies_and_platform_model} we describe the
structural asymmetry in the operational policies (\cref{sec:operational_policies}) and platform
model (\cref{sec:platform_model}) along with the challenges they represent for the applications in
our focus group. In \cref{sec:recommendations} we recommend a more balanced and symmetric HPC
platform model, based on \textit{isolated, user-driven software environments}
(\cref{sec:rec_deployment_model}), network and filesystem \textit{I/O as schedulable resources}
(\cref{sec:rec_schedulable_io}), and improved \textit{introspection and control flow}
(\cref{sec:rec_information_flow}) between platforms and applications. In \cref{sec:implementation}
we present \chpc, an early prototype implementation of our extended platform model based on
\textit{operating system-level virtualization} and \textit{Linux containers} (\cref{sec:chpc}). We
suggest how such a system can coexists with existing batch queueing systems
(\cref{sec:applicability}). Finally in \cref{sec:discussion}, we list related work
(\cref{sec:related_work}), and lay out our upcoming research agenda (\cref{sec:road_ahead}). The
main contributions of this paper are:

\noindent
\begin{enumerate}

  \setlength{\itemsep}{0pt}
  \setlength{\parskip}{0pt}
  \setlength{\parsep}{0pt}

  \item \textit{Identification of existing friction and issues} in application development and
        between HPC platform providers and their tenants.

  \item Recommendations for a \textit{more symmetric HPC platform model} based on decentralized
        software deployment and symmetric interfaces between platforms and applications.

  \item A \textit{non-disruptive candidate implementation} of the conceptual platform model based
        on operating system-level virtualization and containers that can operate alongside existing
        HPC queueing systems.

  \item A \textit{research agenda and statement} to further explore novel HPC platform models based on
        operating-system level virtualization.

\end{enumerate}

%%%%%%%%%%%%%%%%%%%%%%%%%%%%%%%%%%%%%%%%%%%%%%%%%%%%%%%%%%%%%%%%%%%%%%%%%%%%%%%%%%%%%%%%%%%%%%%%%
%%
\section{Focus Applications}
\label{sec:focus_applications}

The authors have deep experience in architecting, developing and running a diverse portfolio of
second generation high-performance and distributed computing applications, tools and frameworks.
These include tightly-coupled parallel codes; distributed, data-intensive and dynamic applications,
and higher-level application and resource management frameworks. This experience shaped the position
we are taking. This section characterizes the applications and the challenges they are facing on
today's HPC platforms.

It would be false to claim that current production HPC platforms fail to meet the requirements of
their application communities. It would be equally wrong to claim that the existing platform model
is a pervasive problem that generally stalls the innovation and productivity of HPC applications. It
is important to understand that significant classes of applications, often from the monolithic,
tightly-coupled parallel realm, have few concerns regarding the issues outlined in this paper. These
applications produce predictable, static workloads, typically map well to the hardware architectures
and network topologies. They are developed and hand-optimized to utilize resources as efficiently as
possible. They are the original tenants and drivers of HPC and have an effective social and
technical symbiosis with their platform environments.

However, it is equally important to understand that other classes of applications (that we call
second generation applications) and their respective user communities share a less rosy perspective.
These second generation applications are typically non-monolithic, dynamic in terms of their runtime
behavior and resource requirements, or based on higher-level tools and frameworks that manage
compute, data and communication. Some of them actively explore new compute and data handling
paradigms, and operate in a larger, federated context that spans multiple, distributed HPC clusters.
When evaluating the challenges, opportunities and recommendations that are laid out in this  paper,
the reader should keep in mind the following three classes of applications.

%%%%%%%%%%%%%%%%%%%%%%%%%%%%%%%%%%%%%%%%%%%%%%%%%%%%%%%%%%%%%%%%%%%%%%%%%%%%%%%%%%%%%%%%%%%%%%%%%
%%
\subsection{Data-Intensive Applications}
\label{sec:data_intensive_applications}

Data-intensive applications require large volumes of data and devote a large fraction of their
execution time to I/O and manipulation of data. Careful attention to data handling is necessary to
achieve acceptable performance or completion. They are frequently sensitive to local storage for
intermediate results and reference data. It is also sensitive to the data-intensive frameworks and
workflow systems available on the platform and to the proximity of data it uses. This may be as a
result of complex input data, e.g. from many sources, with potentially difficult access patterns,
or with requirements for demanding data update patterns, or simply large volumes of input, output or
intermediate data so that I/O times or data storage resources limit performance.

Examples of large-scale, data-intensive HPC applications are \textit{seismic noise
cross-correlation} and \textit{missfit calculation}  as encountered, e.g. in the VERCE
project~\cite{IEEEVERCEVRE2015-SHORT}. Such computations are the only way of observing the deep earth and
are critical in hazard estimation and responder support. The forward wave simulations using
SPECFEM-3D \cite{5215501-SHORT} impose very demanding loads on today's HPC clusters with fast cores and
high-bandwidth interconnect. Critical geophysics phenomena are 3 orders of magnitude smaller than
current simulations. Hence another factor of $10^9$ in computational power could be used. Inverting
the seismic signals to build earth models of sub-surface phenomena requires iterations that run the
forward model, compare the results with seismic observations, \textit{misfit analysis}, at each
seismic station, and compute an adjunct to back propagate to refine the model. The models are
irregular finite element 3D meshes with $10^6 to 10^7$ cells. The noise correlations are modeled as
complex workflows ingesting multivariate time series from more than 1000 seismic stations. These
data are prepared by a pipeline of pre-processing, analysis, cross-correlation and post-processing
phases.

The main issues we encountered with these applications were the difficulty of establishing an
environment that met all the prerequisites, the difficulty of establishing suitable data proximity,
the difficulty of efficiently handling intermediate data, the difficulty of enabling users to
inspect progress and the difficulty of arranging the appropriate balance of the properties of the
hardware context. Coupling concurrent parts of application running on different platforms (to which
they were well suited), porting the applications to new platforms and avoiding moving large volumes
of data often proved impossible.

%%%%%%%%%%%%%%%%%%%%%%%%%%%%%%%%%%%%%%%%%%%%%%%%%%%%%%%%%%%%%%%%%%%%%%%%%%%%%%%%%%%%%%%%%%%%%%%%%
%%
\subsection{Dynamic Applications}
\label{sec:dynamic_applications}

Dynamic applications fall into two broad categories: (i) applications for which we do not have full
understanding of the runtime behavior and resource requirements prior to execution and (ii)
applications which can change their runtime behavior and resource requirements during execution.
Dynamic applications are driven by adaptive algorithms that can require different resources
depending on their input data and parameters. e.g. a data set can contain a specific area of
interest which triggers in-depth analysis algorithms or a simulation can yield an artifact or
boundary condition that triggers an increase in the algorithmic resolution. Examples of dynamic HPC
applications are: (a) applications that use ensemble Kalman-Filters for data assimilation in
forecasting (e.g.~\cite{jha2008developing}), (b) simulations that use adaptive mesh refinement (AMR)
to refine the accuracy of their solutions (e.g.~\cite{berger1989amr}), and (c) seismic wave
propagation simulations that modify their code by using compilation in their early phases (e.g.
SPECFEM3D~\cite{5215501-SHORT}). Many other examples exist.

The main issues we encountered running dynamic applications on production HPC platforms originate in
their dynamic resource and time requirements. A Kalman-Filter application might run for two
hours or for four hours, depending on the model's \textit{initial conditions}. Similarly an AMR
simulation of a molecular cloud might require an additional 128 CPU cores during its computation to
increase the resolution in an area of interest. Resource managers on production HPC platforms do not
support such requirements: the maximum runtime is restricted by the \textit{walltime limit} set at
startup. Resource requirements, e.g. CPU cores and memory, are similarly set at startup. It is
neither possible to request an extension of the runtime nor inform a resource manager about reducing
or increasing requirements.

This inflexibility of the platforms has lead to interesting, yet obscure application architectures:
in ~\cite{jha2008developing} for example, the application is forced to opportunistically allocate
additional resources via an SSH connection to the job manager on the cluster head node and release
them if they are not required. This technique (that can be found in many other applications) wastes
platform resources and increases the complexity of the applications significantly by adding complex
resource management logic, which detracts from their stability and reliability. Generally,
application developers are very creative when it comes to circumventing platform restrictions.
Overlay resource management systems such as \textit{pilot jobs} are becoming increasingly popular
exactly because they enable applications to achieve this effect. We see problems with this approach
for applications and platforms. It adds active resource management as a burden on the shoulders of
developers and users. It dilutes the focus of the applications and adds more complexity and
additional dependencies on potentially short-lived software tools. It increases the expertise
required to develop dynamic applications and consequently restricts widespread adoption of adaptive
techniques. From the platform's perspective, circumventive methods invariably lead to poorer
platform utilization. On the other hand, dynamic applications without active resource management
lead to \textit{hollow utilization} as applications terminate without producing useful results or
without a recent checkpoint.

%%%%%%%%%%%%%%%%%%%%%%%%%%%%%%%%%%%%%%%%%%%%%%%%%%%%%%%%%%%%%%%%%%%%%%%%%%%%%%%%%%%%%%%%%%%%%%%%%
%%
\subsection{Federated Applications}
\label{sec:ds_research_prototypes}

% TODO for this section:
% - mention apache storm

Federated HPC environments have become more and more prominent in recent years. Based on the idea
that federation fosters collaboration and allows scalability beyond a single platform, policies  and
funding schemes explicitly supporting the development of concepts and technology for HPC federations
have been put into place. Larger federations of HPC platforms are XSEDE in the US, and the PRACE~ in
the EU. Both platforms provide access to several TOP-500 ranked HPC clusters and an array of smaller
and experimental platforms. With policies and federated user management and accounting in place,
application developers and computer science researchers are encouraged to develop new application
models that can harness the federated resource in new and innovative ways. Examples for resource
federation systems are \textit{CometCloud}~\cite{Diaz-Montes:2015:CES:2709096.2709104-SHORT} and
\textit{RADICAL Pilot}~\cite{DBLP:journals/corr/MerzkySTJ15}. RADICAL Pilot is a pilot-job system
that allows transparent job scheduling to multiple HPC platforms via the SSH protocol. CometCloud is
an autonomic computing engine for Cloud and HPC environments. It provides a shared coordination
space via an overlay network and various types of programming paradigms such as Master/Worker,
Workflow, and MapReduce/Hadoop. Both systems have been used to build a number of different federated
computational science and engineering applications, including distributed replica exchange molecular
dynamics, ensemble-based molecular dynamics workflows and medical image analysis. The federation
platform provides the execution primitives and patterns for the applications and marshals the job
execution as well as the data transfer of input, output and intermediate data from, to and in
between the different HPC platforms.

Deployment and application mobility has been the biggest issue  with federated applications and
overlay platform prototypes. Even if federated platform use could be shown to work conceptually, in
practice the applications were highly sensitive to and would often fail because of the software
environment on the individual platforms. Software environments are not synchronized or federated
across platforms. As a result, different versions of domain software tools caused the application to
fail. Automated compiling and installing applications in user-space was difficult and sometime
impossible due to incompatible compilers, runtimes and libraries on the target platform. Maintaining
a large database of tools, versions, paths and command line scripts for the individual platforms was
a significant fraction of the overall development effort. Another issue was the limited resource
allocation, monitoring and control mechanisms provided by the platforms, which would be crucial for
an overarching execution platform to make informed decisions.  Analogous to the discussion
in~\cref{sec:dynamic_applications}, applications are bound to static resource allocation. A common
pattern we observed was that federated application would schedule resources on different platforms
and just use the one that became available first. We further observed that application users were
largely unsuccessful in allocating larger number of resources on multiple platforms concurrently
due to missing resource allocation control. This makes the idea of having very large applications
spanning more than one platform very difficult to achieve with production HPC platforms.

%%%%%%%%%%%%%%%%%%%%%%%%%%%%%%%%%%%%%%%%%%%%%%%%%%%%%%%%%%%%%%%%%%%%%%%%%%%%%%%%%%%%%%%%%%%%%%%%%
%%
\section{Challenges and Opportunities}
\label{sec:policies_and_platform_model}

In this section we describe the details of operational policies and properties of the dominant HPC
platform model that we have identified as structurally hindering in our every day work with
second generation HPC applications. The platform model describes the underlying model and
abstractions of the software system that interfaces an HPC cluster with its users. It defines the
views and the interfaces users and application have of the system. Important aspects of the platform
model are (1) the interfaces provided to execute the application on the platform, (2) the view of
the platform's hardware resources while application is running, (3) the view of the application
while it is running on the HPC platform, (4) interfaces provided to control the application while
running on the platform.

The platform model is determined by the software system that is used to manage the platform. As the
\textit{dominant} platform model, we identify HPC platforms that are (1) managed by a job manager  /
queueing system, (2) provide a shared file system across all platform nodes, (3) use the concept of
\textit{jobs} as the abstraction for executing applications, and (4) which provide a single, global
application execution context, the host operating system execution context. We consider it
\textit{dominant} because all production HPC platforms we have worked on and are aware of exhibit
the same model. Only in the implementation details we observed differences between platform, e.g. in
their choice of distributed file systems, queueing systems (PBS, SLURM, LoadLeveler, etc.) and host
operating systems.

%%%%%%%%%%%%%%%%%%%%%%%%%%%%%%%%%%%%%%%%%%%%%%%%%%%%%%%%%%%%%%%%%%%%%%%%%%%%%%%%%%%%%%%%%%%%%%%%%
%%
\subsection{Existing Operational Policies}
\label{sec:operational_policies}

%%%%%%%%%%%%%%%%%%%%%%%%%%%%%%%%%%%%%%%%%%%%%%%%%%%%%%%%%%%%%%%%%%%%%%%%%%%%%%%%%%%%%%%%%%%%%%%%%
%%
\subsubsection*{Software Provisioning}

Software provisioning is a major issue that we have observed throughout all classes and types of HPC
applications, not just the focus applications. Resource providers put a significant amount of effort
into curating up-to-date catalogues of the software libraries, tools, compilers and runtimes that
are relevant to their user communities. Software management tools like \textit{SoftEnv} and
\textit{Module} are commonly used to support this task. Because existing HPC platform models do not
provide software environment isolation (like for example virtualized platforms), all changes made to
a platform's software catalog have an impact on all users. Hence, the process is strongly guarded by
the resource providers. Versioning and compatibility of individual software packages need to be
considered with every update or addition to the catalog. As a consequence, getting an application
and/or its dependencies installed on a platform requires direct interaction with, and in many cases
persuasion of the resource provider. While some software packages are considered less critical,
others, like for example an alternative compiler version, Python interpreter or experimental MPI
library are considered \textit{disruptive} and deployment is often refused. Software deployment in
user space (i.e., the user's home directory) is an alternative, but in practice it has shown to be
very tedious, error prone and difficult to automate. Affected from the software provisioning dilemma
is also application mobility (migratability). Because software environments cannot be shared
between different platforms, application mobility comes at the cost of a significant deployment
overhead that increases linearly with the number of HPC platforms targeted. In several application
projects we have experienced software provisioning as very hard and a time and resource consuming
process. Especially for projects that aim at federated usage of multiple HPC platforms, software
deployment becomes a highly impeding factor.

\subsubsection*{Networking}

In- and outbound networking differs between platforms. Limitations are determined by the platform
architecture and configuration, as well as the networking policies (firewalls) of the organization
operating the platform. We have encountered everything, from platforms with no restrictions, to
platforms from which in- and outbound network connections were virtually impossible. These
differences made it extremely difficult, not only to federate platforms, but also to migrate
applications. We observed several cases in which applications simply were not able to run on a
specific platform because they were design around the assumption that communication and data
transfer between an HPC platform and the internet is not confined. Affected were applications that
dynamically load data from external servers or databases during execution or that rely on methods
for monitoring and computational steering. In other cases the application's performance was severely
crippled as it had to funnel all through through the head node instead of loading the data into the
compute nodes directly because compute nodes could not ``dial out''. In addition, it is not possible
to query platforms for their networking configuration programmatically. Platform documentation or
support mailing-lists are often the only way to gather this information.

\subsection{Existing Platform Model}
\label{sec:platform_model}

\subsubsection*{Static Resource Mapping}
\label{sec:static_resource_mapping}

%%%%%%%%%%%%%%%%%%%%%%%%%%%%%%%%%%%%%%%%%%%%%%%%%%%%%%%%%%%%%%%%%%%%%%%%%%%%%%%%%%%%%%%%%%%%%%%%%
%%
% \subsubsection*{Static Wall-Time Limits}

Existing platform models require users to define an application's expected total runtime, CPU and
memory requirements prior to its execution. None of the job managers found in production HPC
platforms deviates from this model or allows for an amendment of the expected total runtime during
application execution. In our experience, enforcing static wall-time limits lead to two unfavorable
scenarios. In the first scenario, applications run at risk of being terminated prematurely because
their wall-time limit was set to optimistically. Especially dynamic applications are affected by
this. For the users this means that they wasted valuable resource credits without producing any
results. For the platform provider this means \textit{hollow utilization}. In the second scenario,
users ``learn'' from the first scenario and define the application wall-time limit very
pessimistically. Most job managers weigh application scheduling priority against requested resources
and wall-time. The higher the requested wall-time limit, the longer an application has to wait for
its slot to run. This can significantly decrease user productivity. If the application finishes
ahead of its requested wall-time limit, the platform's schedule is affected, which can result in
suboptimal resource utilization. The same limitations hold true for CPU and memory resources with
similar implication for the applications. Especially for dynamic applications, static resource
limits can become an obstacle to productivity and throughput (see also
\cref{sec:dynamic_applications}).

%%%%%%%%%%%%%%%%%%%%%%%%%%%%%%%%%%%%%%%%%%%%%%%%%%%%%%%%%%%%%%%%%%%%%%%%%%%%%%%%%%%%%%%%%%%%%%%%%
%%
\subsubsection*{Asymmetric Resource Model}
\label{sec:asymmetric_resource_model}

%%%%%%%%%%%%%%%%%%%%%%%%%%%%%%%%%%%%%%%%%%%%%%%%%%%%%%%%%%%%%%%%%%%%%%%%%%%%%%%%%%%%%%%%%%%%%%%%%
%%
% \subsubsection*{Unpredictable I/O Performance}

While static resource mapping can be a significant obstacle, it also ensures guaranteed resource
availability and exclusive usage. Job managers require the user to define a fixed number of CPU
cores and optionally the required memory per CPU core. The required network I/O bandwidth and
filesystem I/O operations per second (IOPS) however can not be specified. When working with
data-intensive applications and applications that periodically need to read or write large amounts
of data in bursts, this can lead to unforeseeable variations in the overall performance and runtime
of the application. Because network file system I/O bandwidth is shared with all other tenants on a
platform, the I/O bandwidth available to a user's application critically depends on the I/O load
generated by its peers. We have experienced significant fluctuations in application runtime (and
resulting failure) because of unpredictably decreasing I/O bandwidth (see also
\cref{sec:data_intensive_applications}). We have experienced similar issues with network I/O
bandwidth. Just like file system bandwidth, the in- and outbound network connections are shared
among all platform tenants. Bandwidth requirements can not be defined. Depending on the network's
utilization, the available bandwidth available to an application can fluctuate significantly. For
data-intensive applications that download input data and upload output data from and to sources
outside the platform boundaries, this can become a non-negligible slowdown factor and potential
source of failure.

%%%%%%%%%%%%%%%%%%%%%%%%%%%%%%%%%%%%%%%%%%%%%%%%%%%%%%%%%%%%%%%%%%%%%%%%%%%%%%%%%%%%%%%%%%%%%%%%%
%%
\subsubsection*{Missing Introspection, Control and Communication}
\label{sec:information_flow}

Applications evolve over time through an iterative loop of analysis and optimization. Every time a
new algorithm or execution strategy is added or a new platform is encountered, its implications on
the performance and stability of the application needs to be evaluated. For that, it is critical to
understand the behavior of the application processes their interaction with the platform and their
resource utilization profile. Similarly, these insights are crucial for federated applications and
frameworks (see also~\cref{sec:ds_research_prototypes}) to make autonomous decisions about
application scheduling and placement. HPC platforms currently provide very few tools that can help
to gain these insights and few are integrated with the platform. Interfaces to extract operational
metrics of the resources and the application's jobs and processes are almost entirely missing on
production HPC platforms.

The same limitations hold true for the communication channels between platforms and applications.
The controls from application to platform are effectively limited to starting and stopping user
jobs. The interface is usually confined to the job manager's command line tools and not designed for
programmatic interaction. In the opposite direction, communication is confined to operating system
signals emitted by the platform. Applications can then decide whether they want to implement signal
handlers to react to the signals dispatched by the platform or not. Signals are little more than
notifications about imminent termination.

%%%%%%%%%%%%%%%%%%%%%%%%%%%%%%%%%%%%%%%%%%%%%%%%%%%%%%%%%%%%%%%%%%%%%%%%%%%%%%%%%%%%%%%%%%%%%%%%%
%%
\section{Recommendations}
\label{sec:recommendations}

We have identified static resource mapping, an incomplete resource model, missing introspection and
control, and a centralized, inflexible software deployment model as the main inhibitors to second
generation HPC applications. In this section we lay out the blueprint for a more balanced platform
model that incorporates introspection, bidirectional information and control flow and decentralized
deployment as first-order building blocks.

%%%%%%%%%%%%%%%%%%%%%%%%%%%%%%%%%%%%%%%%%%%%%%%%%%%%%%%%%%%%%%%%%%%%%%%%%%%%%%%%%%%%%%%%%%%%%%%%%
%%
\subsection{Introspection and Control Model}
\label{sec:rec_information_flow}

Many HPC applications and higher-level application frameworks do not implement common resilience and
optimization strategies even though the knowledge is available  via research publications and
prototype systems. We identify the need for interfaces to retrieve the physical and logical model,
state of an application, the physical model and state of the platform. This addresses the asymmetric
platform issue, where once an application is \textit{submitted} to an HPC platform,  users loose
insight and control over their application almost entirely. Existing point solutions to establish
control and introspection for running applications are often difficult to adapt or their deployment
is infeasible due to their invasiveness. Furthermore, introspection implemented redundantly on
application-level adds additional pressure on a platform's resources and adds additional sources of
potential error. Based on these premises, we propose a platform model that incorporates and builds
upon \textit{symmetric} introspection, information- and control-flow across the
platform-application-barrier.

\subsubsection*{Physical and Logical Application Models}
\label{sec:physical_and_logical_application_models}

Our approach distinguishes between the \textit{logical} and the \textit{physical} model and state of
an application. The logical model and state is rendered within the application logic and is designed
by the application developer. Logical state consists e.g. of the current state of an
algorithm, the current state / progress of the simulation of a physical model, or whether an
application is e.g. in startup, shutdown or checkpointing state. The logical model is
inherently intrinsic to the application and can only be observed and interpreted fully within its
confines. In contrast, the physical model and state of an application captures the executing
entities that comprise a running application: OS processes and threads. Process state (also called
\textit{context}), informs about the locality, resource access and utilization of a process. This
includes memory consumption, network and filesystem I/O. Together, the physical and the logical
state make up the overall state of an application.

Intuitively, logical state determines the physical state of an application. However, changes in the
physical state (e.g. failure) can influence the logical state as well. The third important aspect
is the \textit{platform} state, which also influences the physical state of an application (e.g.
resource throttling, shutting-down). This means that we end up with two entities that influence the
physical state of the application with potentially contradictory goals: the state of the platform
and the application model.

% Based on our assumption in section \ref{sec:intro} that applications and platform providers  have a
% common mission, we posit that it is hence crucial to provide interfaces that allow introspection of
% the physical application state, aspects of the logical application state and the platform state.
%
\subsubsection*{Physical Application Model and Interface}
\label{sec:physical_application_model}

A physical application model should be able to capture the current state and properties of the
executable entities of an application with respect to their relevance to resilience and optimization
strategies of the application and the platform. A physical state interface should allow real-time
extraction of the physical state and properties. As the physical state of an application is the
state of its comprising processes (and threads), the interface should provide at least the following
information (1) application processes and their locality (placement), (2) memory consumption, (3)
CPU consumption, (4) filesystem I/O activity, (5) filesystem disk usage, (6) network activity
(platform in-/outbound), and (7) inter-process network activity.

\begin{figure}[t]
  \centering
    \includegraphics[width=\columnwidth]{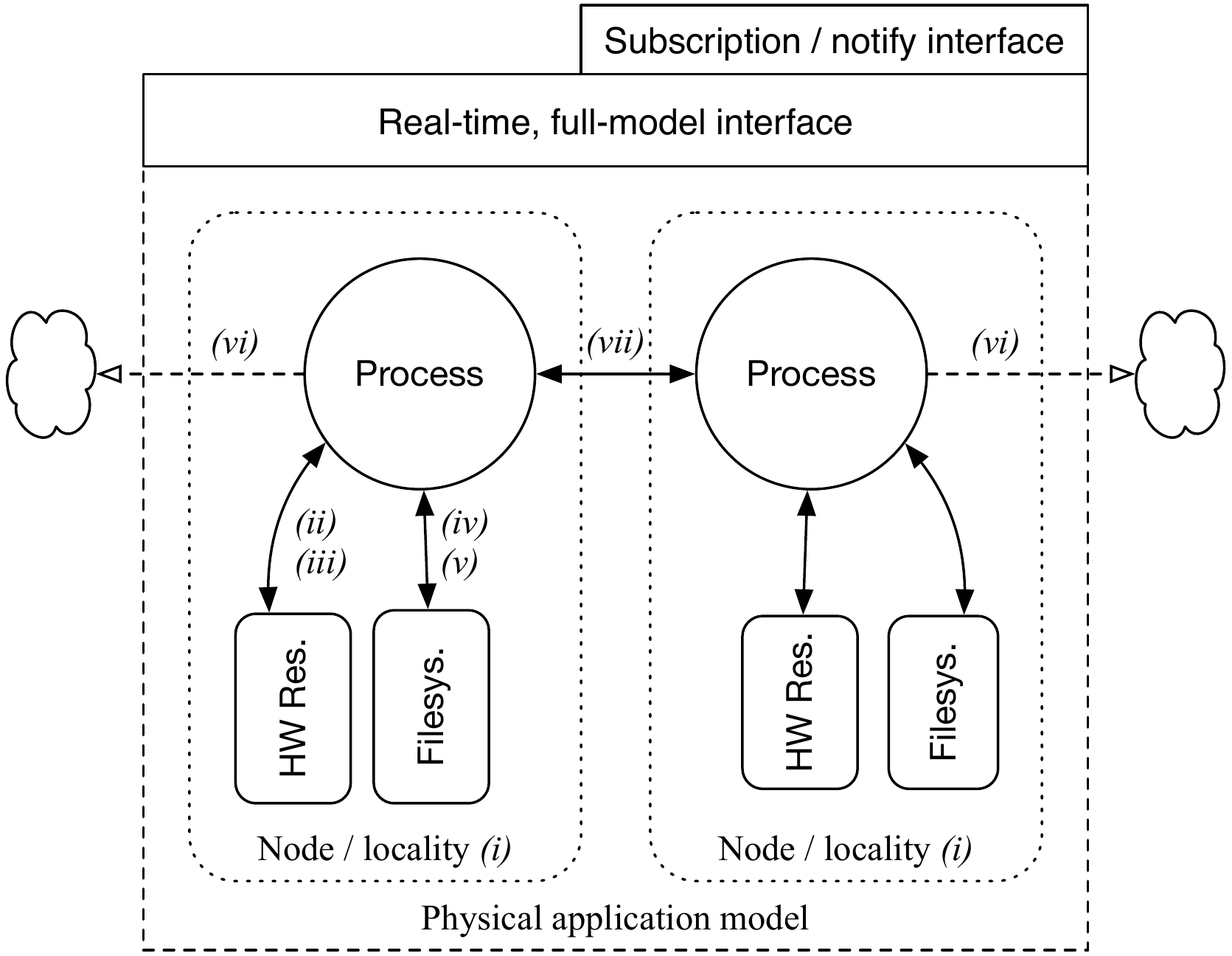}
    \caption{Physical application model. Changes in logical application state and
     platform state can be observed within this model.}
  \label{fig:physical_model}
  \vspace{-1.1em}

\end{figure}

With this information available, a complete, real-time physical model of the application can be
drawn (see Figure ~\ref{fig:physical_model}). The results of changes to the logical application
state and the platform state can be observed within this model, hence it can serve as the foundation
for both manual and automated application analysis and optimization, bottleneck and error detection.

Real-time physical state information is not always trivial to handle and to interpret, especially
not at large scales. While having the complete physical state information of an application
available is crucial for advanced use-cases, it can turn out to be impractical or too complex to
handle and analyze for more basic use-cases and applications. Hence, we recommend that a physical
state interface provides a \textit{subscription-based} interface complementary to the real-time,
full-model interface. A subscription-based interface allows boundary conditions to be set for
entities in the physical model. If these boundary conditions are violated, a notification is sent to
the application. e.g. an application might set a boundary condition for maximum memory consumption
or minimum filesystem throughput and act accordingly if any of these boundary conditions is
violated, e.g. by adjusting the simulation models or algorithms.

To allow a maximum degree of flexibility, the physical state interface of an HPC platform should not
be confined to the application itself, but should also be accessible by a higher-level application
frameworks and services, and by human operators through interactive web portals and analysis and
optimization tools.

\subsubsection*{Logical Application Model and Interface}

While the physical state model of an application is explicitly determined by the state of its
individual operating system processes, the logical state model of an application is a lot more
fuzzy, as it is largely defined by the application itself. The majority of logical application
states will not be relevant or parseable by entities outside the application. Hence, we recommend an
implementation of an extensible logical state interface that captures application states that are
(a) relevant outside the application logic, and (b) generic enough to be applicable to a large
number of different HPC applications. The logical application model interface is important as it
allows state information to flow from applications to the platform and its management components,
like schedulers, and accounting services. As a first approximation, we recognize the following
logical application states

\begin{enumerate}
  \setlength{\itemsep}{0pt}
  \setlength{\parskip}{0pt}
  \setlength{\parsep}{0pt}

  \item{\textit{Running}:       executing normally }
  \item{\textit{Checkpointing}: checkpointing current state}
  \item{\textit{Restoring}:     restoring from a checkpoint }
  \item{\textit{Idle}:          waiting for an external event }
  \item{\textit{Error}:         in a terminal state of failure }

\end{enumerate}

\noindent In addition to the application \textit{states}, the the logical application model should
have an optional notion of an application's relative \textit{progress}. With application state and
progress information it is possible to make  basic assumptions about the internal state of the
application. It can help the platform to (a) track the progress of an application, (b) determine a
preferable time for application interruption (e.g. after checkpoints), and (c) make decisions about
resource (re-)assignment and QoS by observing I/O-, compute-centric and idle states.

% Logical state information is provided \textit{voluntarily} by the application via the logical state
% interface of a platform. From the application developer and user perspective, the added development
% costs by integrating state dissemination come with the benefit that suitable platforms will use the
% information to provide an execution environment that is as non-disruptive and resource efficient as
% possible.

\begin{figure}[t!]
  \centering
    \includegraphics[width=\columnwidth]{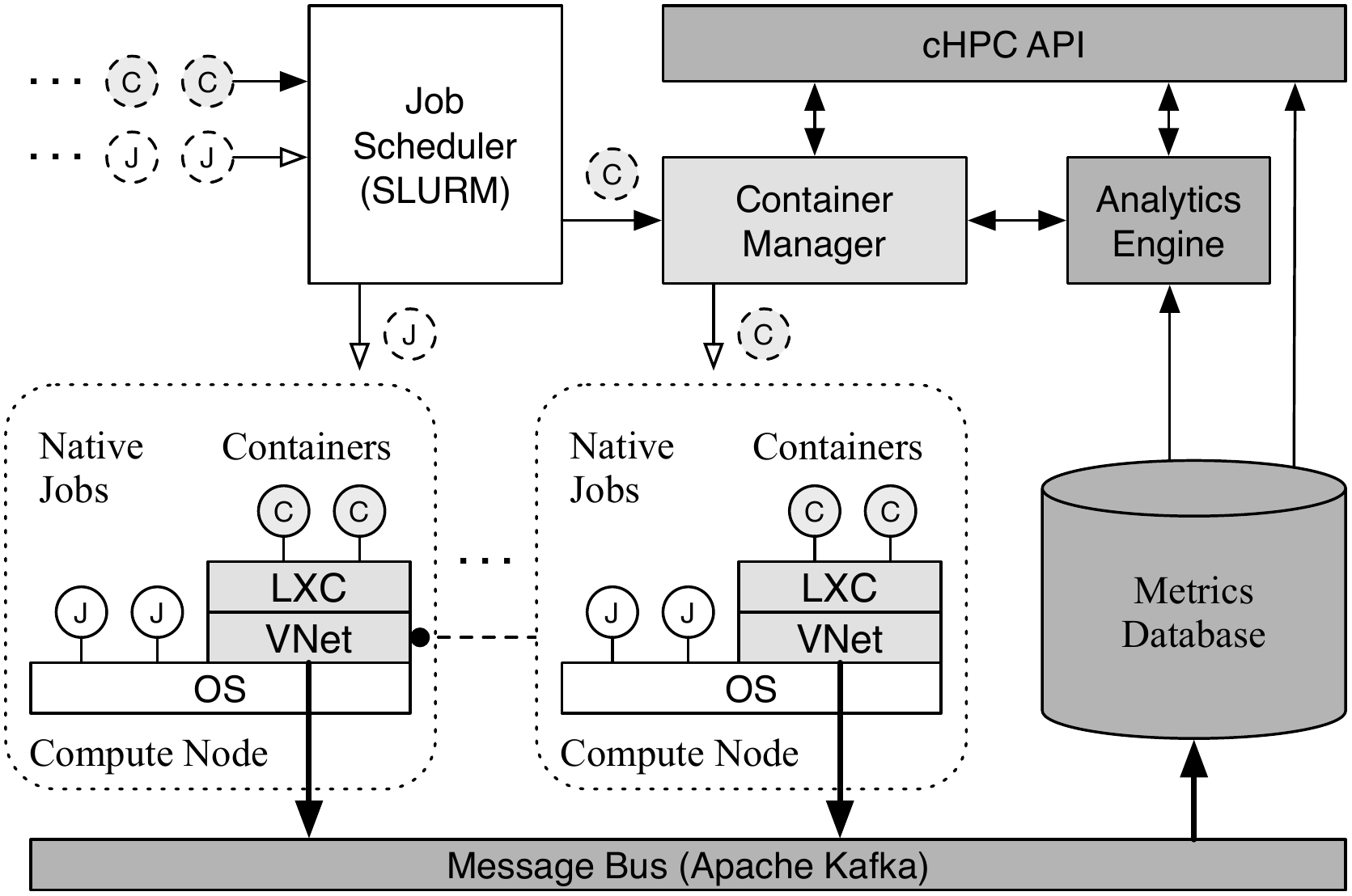}
    \caption{The \chpc{} architecture combines existing HPC platforms with LXC and platform analytics.}
  \label{fig:implementation}
  \vspace{-1.1em}
\end{figure}

\subsubsection*{Platform Environment Model and Interface}

The third and last sub-model that comprises our concept of a more symmetric HPC platform model is
the platform environment model. Analogous to the logical and physical application models, the
platform environment model captures the state and properties of the platform resources (hardware)
and management services (schedulers, QoS) with respect to their relevance for implementing
resilience and optimization mechanisms. The accompanying interface should again allow pull-based
real-time, full-model extraction as well as push-based subscription / notify access.

The platform environment model interface is relevant for the platform management software as well as
applications. For the platform, it provides self-introspection, for the application it provides
environmental awareness. As a minimum, we recommend to capture the following per-node utilization
metrics: (1) CPU, (2) memory, (3) network I/O bandwidth, (4) filesystem I/O bandwidth, and (5)
storage. Platform environment states reflect the platform state with respect to the application:

\begin{enumerate}
  \setlength{\itemsep}{0pt}
  \setlength{\parskip}{0pt}
  \setlength{\parsep}{0pt}

  \item{\textit{Draining}: resources are drained to prepare application termination.}
  \item{\textit{Terminating}: the application is being terminated.}
  \item{\textit{Adjusting}: resources are temporarily reduced or increased}
  \item{\textit{Freezing}: application resources are temporarily withdrawn.}

\end{enumerate}

% %%%%%%%%%%%%%%%%%%%%%%%%%%%%%%%%%%%%%%%%%%%%%%%%%%%%%%%%%%%%%%%%%%%%%%%%%%%%%%%%%%%%%%%%%%%%%%%%%
% %%
% \subsection{Bidirectional Control Flow}
% \label{sec:rec_control_flow}
%
% Introspection and information flow need to be paired with control mechanisms to become
% useful for applications. In \Cref{sec:control_flow} we describe the very limited control flow model
% currently deployed job scheduling systems provide. As part of our platform model, we propose a
% richer control model that works in both directions: from application to platform and from platform
% to application.
%
% \subsubsection{Application to Platform}
% \label{sec:application_to_platform}
%
%
% \subsubsection{Platform to Application}
% \label{sec:platform_to_application}
%
% In order not to disrupt existing applications and usage modes, platform to application control flow
% has to be an optional mechanism. Requiring applications to expose and adhere to a control interface
% in order to run on a specific platform would not be compatible with the idea of an incremental
% platform model. Starting from the existing optional communication channels (see
% \Cref{sec:control_flow}) in the existing platform model, we recommend an extension of the
% communication protocol to better application requirements.

%%%%%%%%%%%%%%%%%%%%%%%%%%%%%%%%%%%%%%%%%%%%%%%%%%%%%%%%%%%%%%%%%%%%%%%%%%%%%%%%%%%%%%%%%%%%%%%%%
%%
\subsection{I/O Resource and Storage Scheduling }
\label{sec:rec_schedulable_io}

In order to address the requirements for a predictable and stable execution environment for
applications that work with large and dynamic datasets, we recommend to make I/O and storage first
order resources in future HPC platform models. We suggest that network and filesystem bandwidth, as
well as storage capacity become reservable entities, just like CPU and memory. It should be possible
for an application to reserve a specific amount of storage space and a guaranteed filesystem and
inbound/outbound network I/O bandwidth. Schedulers would take these additional requests into
account.

%%%%%%%%%%%%%%%%%%%%%%%%%%%%%%%%%%%%%%%%%%%%%%%%%%%%%%%%%%%%%%%%%%%%%%%%%%%%%%%%%%%%%%%%%%%%%%%%%
%%
\subsection{Isolated, User-Driven Deployment}
\label{sec:rec_deployment_model}

Lastly, to address the issues of software and application deployment and mobility (migratability),
we recommend that future HPC platform models embrace a user-driven software deployment approach and
isolated software environments to provide a more hospitable environment for second generation HPC
applications, and equally important, a sustainable environment for legacy applications. Legacy
applications can be equally affected when they slowly grow incompatible with centrally managed
libraries and compilers.

%%%%%%%%%%%%%%%%%%%%%%%%%%%%%%%%%%%%%%%%%%%%%%%%%%%%%%%%%%%%%%%%%%%%%%%%%%%%%%%%%%%%%%%%%%%%%%%%%
%%
\section{Implementation}
\label{sec:implementation}

In this section we provide a brief outline of \chpc{} (\textit{container} HPC), our early prototype
implementation of an HPC platform architecture driven by the recommendations in
\cref{sec:recommendations}. We give a high-level overview of its architecture and implementation and
discuss how it \textit{complements} existing platform models, i.e., allows a non-disruptive,
incremental evolution of production HPC platforms.

\subsection{The cHPC Platform }
\label{sec:chpc}

To explore the implementation options for our new platform model, we have developed \chpc{}, a set
of operating-system level services and APIs that can run alongside and integrate with existing job
% manager-based HPC platforms (see \Cref{fig:implementation}). \chpc{} uses OS-level virtualization
via Linux containers (LXC) to provide isolated, user-deployed application environment containers,
application introspection and resource throttling via the \textit{cgroups} kernel extension. The LXC
runtime and software-defined networking are provided by Docker~\footnote{\scriptsize Docker:
https://www.docker.com} and run as OS services on the compute nodes. Applications are submitted via
the platform's ``native'' job scheduler (in our case \textsf{SLURM}) either as regular HPC jobs that
are launched as processes on the cluster nodes, or as containers, that run supervised by LXC. This
architecture allows HPC jobs and container applications to run side-by-side on the same platform,
which allows direct comparison of the performance and overhead.

To provide platform and application introspection (see ~\Cref{sec:rec_information_flow}), container
and node metrics are collected in real time via a high-throughput, low-latency message broker based
on Apache Kafka~\footnote{\scriptsize Apache Kafka: https://kafka.apache.org} and streamed via the
platform API service to one or more consumers. These can be a user, a monitoring dashboard or the
application itself. The data is also ingested into a \textit{metric database} for further processing
and deferred retrieval. The purpose of the \textit{analytics engine} is, to compare the stream of
platform data with the thresholds set by the applications (\Cref{sec:physical_application_model})
through the platform API and to send an alarm signal to all subscribers when it is violated.

A version of \chpc{} has been deployed on a virtual $128$-core \textsf{SLURM} cluster on $8$
dedicated Amazon EC2 instances. We are in the process of deploying \chpc{} on
EPCC's~\footnote{\scriptsize Edinburgh Parallel Computing Centre: \url{https://www.epcc.ed.ac.uk}}
$24$ node, $1536$-core \textit{INDY} cluster for further evaluation.

\subsection{Practical Applicability}
\label{sec:applicability}

We believe that it is important to find a \textit{practical} and \textit{applicable} way forward
when suggesting any changes to the HPC platform model. Users and platform providers should benefit
from it alike and it should not disrupt existing applications and usage modes. Our implementation
blueprint fulfills these requirements. It is designed to be explicitly non-disruptive, the most
critical aspect for its real-world applicability. Suggesting a solution that would disrupt the
existing platform and application ecosystem would be, even though conceptually valuable, not
relevant for real-world scenarios. The key asset is the non-invasiveness of Docker's
operating-system virtualization. It is designed as an operating-system service accompanied by a set
of virtual devices. Assuming a recent version of the operating system kernel, it can be installed on
the majority of Linux distributions. Additional rules or plug-ins can be added without disruption to
existing job managers to allow them to launch container applications through regular job
descriptions files. The remaining components, introspection API services, metrics database and
analytics engine can run on external utility nodes.

This setup allows us to run containerized applications alongside regular HPC jobs on the same
platform. This is suboptimal from perspective of adaptive resource management and global platform
optimization as regular HPC jobs cannot be considered for optimization by our system. However, it is
a viable way forward towards a possible next incremental step in HPC platform evolution. It also
achieves our main goal, provide a more suitable, alternative  platform for applications that have
difficulties existing within the current HPC platform model and policies.

% \section{Discussion and Future Work}
% \label{sec:discussion}

% \subsection{Related Work}
% \label{sec:related_work}
%
% As opposed to research on HPC architectures and programming models, the research on HPC platform
% models appears as rather sparse. However recent research explores the applicability of Linux
% containers in an HPC context, for example in ~\cite{xavier2013performance},
% ~\cite{felter2015updated-SHORT} and ~\cite{containthis}. The need for application and resource
% monitoring is picked up in many application-centric publications, but e.g.
% \cite{Sharifi:2015:MHA:2726935.2726944} discusses it also directly in an HPC context. Datacenter
% platforms like UC Berkeley's work on what is now Apache Mesos~\cite{hindman2011mesos-SHORT} and
% Google's work on \textit{Borg}~\cite{43438-SHORT} exhibit many of the characteristics that we
% postulate in \cref{sec:recommendations}. They are however disruptive in nature and don't allow for
% an evolution of existing platforms.

\section{The Road Ahead}
\label{sec:road_ahead}

Many of the platform model issues we discuss in this paper are based on our own experience as well
as the experience of our immediate colleagues and collaborators. Also platform providers seem to
have a tightened awareness of these issues. In order to qualify and quantify our assumptions, we are
in the process of designing a survey that will be sent out to platform providers and application
groups to verify current issues on a broader and larger scale. The main focus of our work will be on
the further evaluation of our prototype system. We are working on a ``bare metal'' deployment on HPC
cluster hardware at EPCC. This will allow us to carry out detailed measurements and benchmarks to
analyze the overhead and scalability of our approach. We will also engage with computational science
groups  working on second generation applications to explore their real-life application in the
context of \chpc.

%ACKNOWLEDGMENTS are optional
% \section{Acknowledgments} This research is supported by an \textit{AWS in Education Research} grant
% from Amazon Web Services, Inc., and the Edinburgh Parallel Computing Centre (EPCC).

\bibliographystyle{abbrv}
\bibliography{pospaper-preprint}

\end{document}